%% file: CLIPPER.tex
\documentclass[conference]{IEEEtran}
\IEEEoverridecommandlockouts
\usepackage{cite}
\usepackage{amsmath,amssymb,amsfonts}
\usepackage{algorithmic}
\usepackage{graphicx}
\usepackage{textcomp}
\usepackage{xcolor}
\usepackage{colortbl}
\def\BibTeX{{\rm B\kern-.05em{\sc i\kern-.025em b}\kern-.08em
    T\kern-.1667em\lower.7ex\hbox{E}\kern-.125emX}}

\usepackage{multirow}
\usepackage{makecell}
\usepackage{amsmath}
\usepackage{amssymb}
\usepackage{pifont}
\usepackage[hyphens]{url}
\usepackage{hyperref}
\usepackage{algorithm}
\usepackage{algorithmic}

\begin{document}

\title{Is One-Shot In-Context Learning Helpful for Data Selection in Task-Specific Fine-Tuning of Multimodal LLMs?\\
}

\author{
    Xiao An$^{1}$,~
    Jiaxing Sun$^{2}$,~
    Ting Hu$^{3}$,~
    Wei He$^{1\dag}$ \\
    $^1$Wuhan University~
    $^2$Shanghai AI Laboratory \\
    $^3$Nanjing University of Information Science and Technology \\
    \text{\{anxiao, weihe1990\}@whu.edu.cn, sunjiaxing@pjlab.org.cn, hutingrs@nuist.edu.cn}
    \thanks{$^\dag$Corresponding author.}
}


\maketitle

\input{sec/0_abstract}
\input{sec/1_introduction}
\input{sec/2_related_works}
\input{sec/3_methodology}
\input{sec/4_experiments}
\input{sec/5_conclusion}

\section*{Acknowledgment}
This work was supported by the National Natural Science Foundation of China under Grant 42271370.

\bibliographystyle{IEEEbib}
\bibliography{ref}

\input{sec/X_appendix}

\end{document}

%% file: sec/0_abstract.tex
\begin{abstract}
Injecting world knowledge into pretrained multimodal large language models (MLLMs) is essential for domain-specific applications. Task-specific fine-tuning achieves this by tailoring MLLMs to high-quality in-domain data but encounters scalability challenges as datasets grow, necessitating a trade-off between performance and computational overhead. Existing data selection methods rely on additional scoring models or heuristic clustering, failing to concentrate on both data importance and diversity. Moreover, both methods overlook the interplay among training samples. To address these limitations, we propose CLIPPER, a training-free data selection pipeline that separates parameter and world knowledge, and leverages in-context learning to probe model responses to different demonstration-query combinations. CLIPPER identifies coresets that mirror the original dataset's perplexity distribution, preserving critical samples while maintaining diversity. Experiments on two MLLMs and three datasets show that CLIPPER matches full fine-tuning performance with significantly lower costs: Qwen2.5-VL-7B attains 47\% data efficiency on VRSBench, and Llama-3.2-11B-Vision-Instruct reduces ScienceQA training time by 37\%.
\end{abstract}

\begin{IEEEkeywords}
Multimodal LLM, Data Selection, In-Context Learning, Task-Specific Fine-Tuning
\end{IEEEkeywords}

%% file: sec/1_introduction.tex
\section{Introduction}
\label{sec:introduction}

\begin{figure}[!t]
    \centering
    \includegraphics[width=0.9\linewidth]{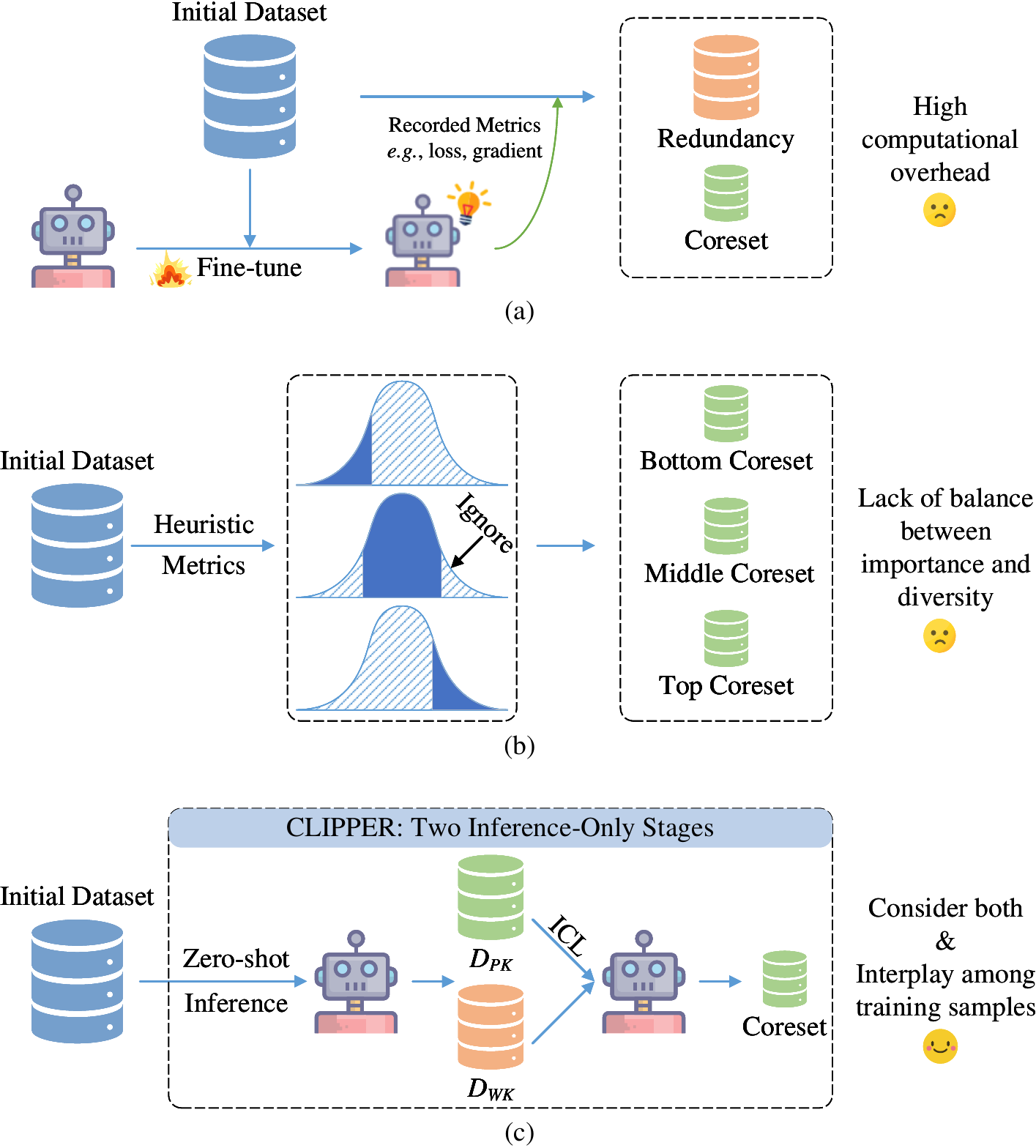}
    \caption{Method comparisons of (a) importance-based, (b) diversity-based, and (c) our CLIPPER. CLIPPER requires only a two-stage inference, eliminating the need for additional fine-tuning and heuristic clustering.} 
    \label{fig1:comparison}
    \vspace{-1.6\baselineskip}
\end{figure}

Building upon the success of large language models (LLMs), general-domain multimodal large language models (MLLMs), such as Qwen2.5-VL~\cite{qwen2.5-vl} and Llama3.2~\cite{llama-3.2}, have reformed conventional visual interaction, exhibiting impressive potential in comprehending and reasoning over visual-linguistic information. To further adapt MLLMs to domain-specific applications, injecting new world knowledge through fine-tuning on domain-specific instruction-following datasets has become the standard paradigm for transferring behavioral norms~\cite{llm-sft-review,sft-review,self-align}. However, this process typically demands substantial training data and computational resources. Prior work~\cite{icl,icl_1,icl_2} has shown that unimodal LLMs can learn from demonstrations through in-context learning (ICL) without modifying parameter knowledge (i.e., without updating model weights). Yet, recent studies indicate that, despite its simplicity and flexibility, ICL still underperforms compared to instruction fine-tuning~\cite{icl-if_1}, which often generalizes better~\cite{icl-if_2}. Therefore, fine-tuning remains the preferred approach for downstream applications.

Expanding task-specific fine-tuning (TSF) datasets to enhance domain-specific proficiency often incurs substantial computational costs and increased carbon emissions, challenging both economic efficiency and environmental sustainability~\cite{llmcarbon}. To mitigate these issues, two main strategies have been proposed for data pruning and optimal coreset selection, as illustrated in Fig.~\ref{fig1:comparison}. However, several challenges persist: \textbf{(1) High Computational Overhead:} importance-based approaches typically involve fine-tuning an additional scoring model—often mirroring the target model—on the datasets for a few epochs to estimate the difficulty of world knowledge embedded in each sample and retain the most important and challenging samples~\cite{importance-1,importance-2,importance-3}. However, this extra training can be even more time-consuming than full fine-tuning itself~\cite{data-whisperer}. \textbf{(2) Lack of Balance Between Importance and Diversity:} diversity-based approaches cluster samples using heuristic metrics based on the parameter knowledge of target models and select subsets to maximize feature-space coverage~\cite{perplexity,diversity-1,diversity-2}. However, relying on a single metric can overlook representative samples in low-scoring regions, yielding biased coresets~\cite{coinside}. \textbf{(3) Underexplored Parameter–World Knowledge Interplay:} the relationship between the intrinsic parameter knowledge of MLLMs and the knowledge embedded in TSF datasets remains insufficiently studied. Additionally, the complex interplay among training samples is generally overlooked, despite their demonstrated importance~\cite{self-align,ft-hallu} and as shown in Fig.~\ref{fig2:fraction}.

The combination of demonstration and new query in ICL, where demonstrations are part of the input, provides a natural setting to explore the interplay between parameter knowledge and world knowledge. Building on this insight, we propose CLIPPER, a training-free pipeline that takes advantage of MLLMs' parameter knowledge and employs ICL to probe the necessity of training samples, enabling optimal coreset selection for subsequent TSF. Specifically, CLIPPER consists of two stages as illustrated in Fig.~\ref{fig3:overview}:

\noindent\textbf{\ding{192} Categorization of Parameter and World Knowledge:} Perform zero-shot inference with the original MLLM and categorize training samples based on the semantic alignment between ground-truth label and model response.

\noindent\textbf{\ding{193} In-Context Learning Probing and Selection:} Use samples identified as parameter knowledge as demonstrations and randomly choose $R$ samples grouped into world knowledge to serve as queries in $R$ one-shot ICL setup. According to the results, original dataset is partitioned into four subsets, whose dynamic combinations can then be used to construct an optimal coreset for downstream TSF.

Such design of CLIPPER utilizes two straightforward stages of naive zero-shot inference and ICL, which can be easily accelerated~\cite{vllm,lmdeploy}, eliminating the need to train additional models or reliance on a single heuristic metric. Further analysis of statistical perplexity across different subsets from both stages reveals that the selected coresets closely mirror the distribution of the original parameter knowledge and world knowledge. This key observation indicates that ICL-driven interplay enables perplexity-aware data selection, emphasizing important training data while maintaining data diversity spontaneously, without explicitly relying on heuristic metric.

To verify the effectiveness and efficiency of CLIPPER in data selection, we compare models fine-tuned on coreset selected by CLIPPER with the fully fine-tuned counterparts across two mainstream MLLMs and three representative downstream tasks (remote sensing, science question answering, and outside-knowledge question answering). Experiment results show that CLIPPER generalizes well across tasks and qualifies the partially fine-tuned models to achieve performance comparable to, or even exceeding, fully fine-tuned ones while reducing considerable training samples and computational overhead. Notably, Qwen2.5-VL-7B attains comparable results on VRSBench despite discarding 39\% of training samples, and Llama-3.2-11B-Vision-Instruct reduces training time by 22\%. In summary, our contributions are as follows:

\begin{itemize}
\item[(1)] We propose CLIPPER, a simple yet effective training-free pipeline for TSF coreset selection, requiring only zero-shot inference followed by ICL, thus avoiding extra scoring model training and improving efficiency.

\item[(2)] We leverage ICL to capture the interplay among training samples, demonstrating that ICL-based probing enables perplexity-aware data selection. The closely aligned perplexity distributions between selected subsets and original datasets suggest that CLIPPER naturally balances data importance and diversity without relying on heuristics.

\item[(3)] Extensive experiments on two representative MLLMs and three diverse downstream datasets demonstrate CLIPPER's efficiency and effectiveness, achieving comparable performance while reducing considerable training time.
\end{itemize}

%% file: sec/2_related_works.tex
\section{Related Works}
\label{sec:related_works}

\subsection{Parameter and World Knowledge in TSF Datasets}
~\cite{self-align} investigated the influence of injecting pure world knowledge or pure parameter knowledge into unimodal LLMs by fine-tuning on four downstream tasks. Their results reveal that TSF data fully aligned with the model’s parameter knowledge does not always yield optimal performance, and large discrepancies between the world knowledge in the TSF data and the model’s existing parameter knowledge can substantially degrade capabilities. Similarly, ~\cite{ft-hallu} further fine-tuned LLMs on varying proportions of parameter and world knowledge, reinforcing these observations. However, such investigations have not yet been extended to multimodal context.

\subsection{In-Context Learning}
In-Context Learning (ICL) is a task adaptation technique that enables pretrained models to perform domain-specific tasks without modifying parameter knowledge~\cite{icl}. ICL conditions models on a sequence of demonstration pairs, each consisting of an input query and its ground-truth label. During inference, the model receives these demonstrations followed by a new query, and predicts the corresponding label guided by the context. Recent works~\cite{icl-if_1,icl-if_2} have examined the relationship between ICL and fine-tuning, concluding that while ICL can match fine-tuning under certain decoding settings, it generally exhibits weaker generalization. Nevertheless, the effectiveness of ICL using inference alone motivates us to explore its potential for improving TSF data selection methods.

\subsection{Data Selection}
As training sets grow, especially for MLLMs, how to strike a balance between performance and computational cost remains a major challenge. Two main data selection strategies have therefore gained popularity: (1) Importance-based methods~\cite{self-align,ft-hallu} train an auxiliary scoring model to identify and select the most important and challenging samples. (2) Diversity-based methods~\cite{perplexity,diversity-1,diversity-2} leverage heuristic metrics to select samples across diverse regions in the feature space. However, the former often incurs considerable training costs, while the latter may overlook representative samples in low-scoring regions. Considering these two limitations and the importance of the interplay among training samples, CLIPPER requires only zero-shot inference followed by corresponding ICL, avoiding extra fine-tuning and eliminating the need for explicit heuristics.

%% file: sec/3_methodology.tex
\section{Methodology}
This section presents the experiment demonstrating the critical role of the interplay between parameter knowledge and world knowledge in MLLMs. Based on this insight, we introduce CLIPPER that leverages ICL for data selection.

\begin{figure}[!t]
    \centering
    \includegraphics[width=0.70\linewidth]{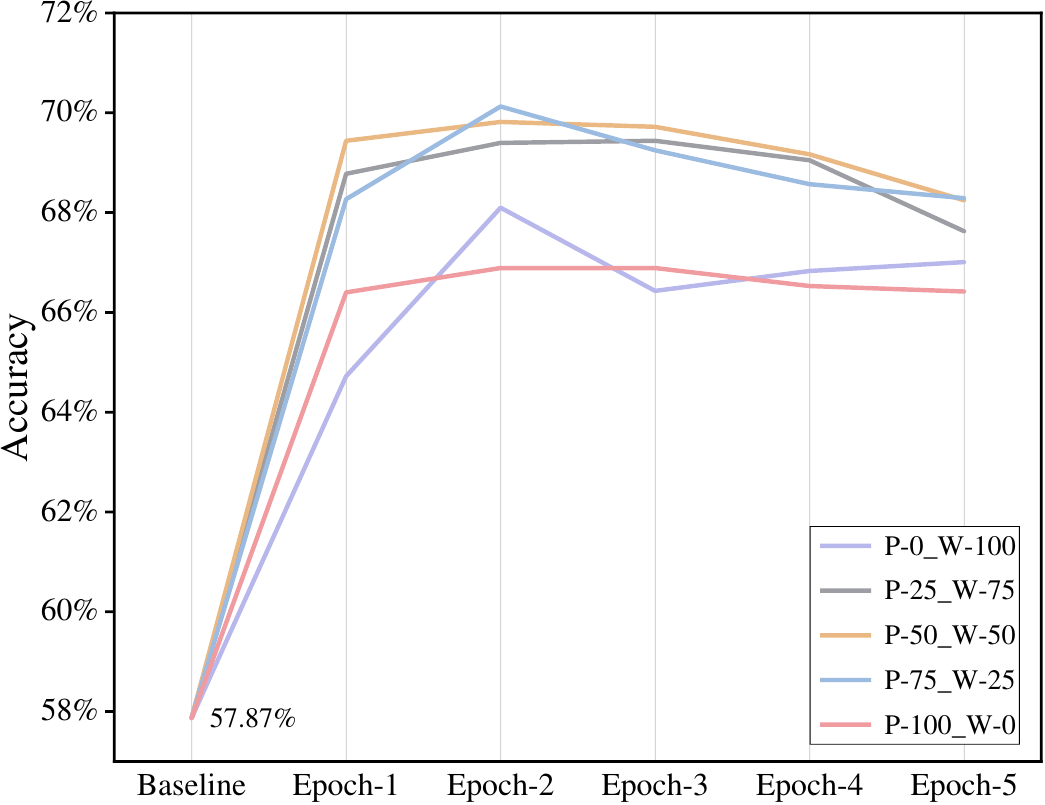}
    \caption{Performance of Qwen2.5-VL-7B on VRSBench. The total training size is fixed and the ratio of parameter- and world-knowledge samples (P-x/W-y) is varied.}
    \label{fig2:fraction}
    \vspace{-1.6\baselineskip}
\end{figure}

\subsection{Interplay among Training Samples}
Training samples are considered part of MLLM's parameter knowledge if its responses align with ground-truth labels; otherwise, they are categorized as new world knowledge~\cite{self-align}. Prior studies on unimodal LLMs have uncovered that improvements on domain-specific tasks arise not from pure parameter or world knowledge alone, but from their interplay~\cite{self-align,ft-hallu}. However, this phenomenon has not been explored in multimodal context. We carefully design an experiment, illustrated in Fig.~\ref{fig2:fraction} (see Appendix for details), showing that the mixture of two types of training samples outperforms using either type alone, while increasing only the proportion of world knowledge does not lead to performance gains. This indicates that the performance degradation of prior data selection methods can be further attributed to their narrow focus on either importance or diversity, while neglecting the crucial interplay between knowledge types. Furthermore, it also reveals substantial redundancy in randomly selected training samples, reinforcing the necessity for data selection strategies that explicitly account for inter-sample relationships.

\begin{figure}[!t]
    \centering
    \includegraphics[width=0.95\linewidth]{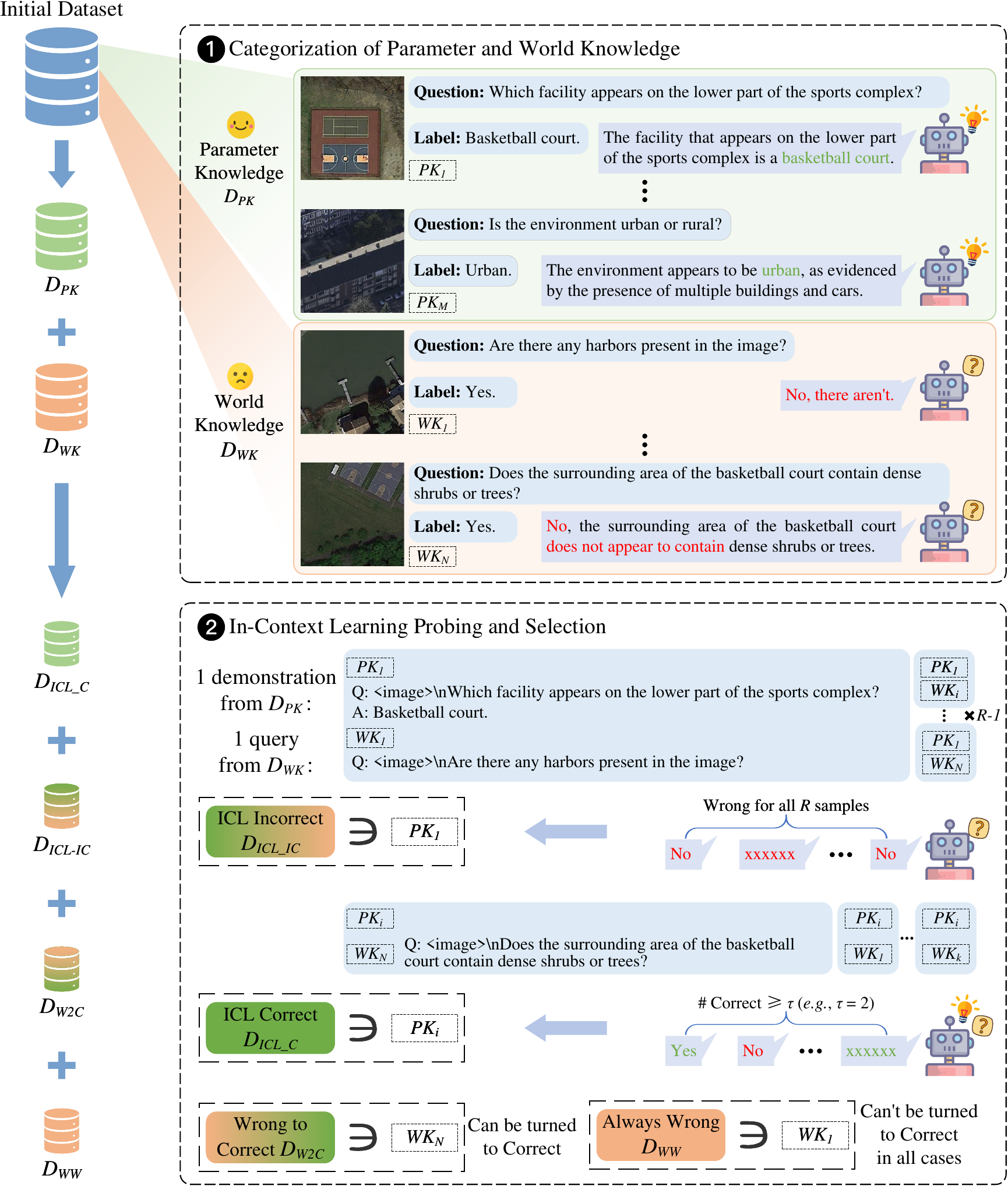}
    \caption{Two-stage inference workflow for CLIPPER data selection. The original datasets are divided into four subsets for diverse combinations.} 
    \label{fig3:overview}
    \vspace{-1.6\baselineskip}
\end{figure}

\subsection{CLIPPER}
Given the high computational cost of additional fine-tuning, the limited focus on either data importance or diversity, and the aforementioned findings, we propose CLIPPER, a training-free data selection method that integrates the interplay between parameter and world knowledge by ICL. CLIPPER operates through two inference-only stages, as illustrated in Fig.~\ref{fig3:overview}. In the \textit{Categorization of Parameter and World Knowledge} stage, the target MLLM performs zero-shot inference to distinguish whether training samples reflect existing parameter knowledge or introduce world knowledge. In the \textit{In-Context Learning Probing and Selection} stage, samples from both categories are randomly paired to form one-shot ICL inputs that probe the model’s responses under contextual guidance. Based on the results, four subsets are grouped ultimately for coreset selection.

\noindent\textbf{$\bullet\quad$Stage 1: Categorizing Parameter and World Knowledge}
A portion of the TSF dataset aligns with the intrinsic parameter knowledge of the target MLLM, while the remaining are considered as world knowledge to be injected~\cite{self-align}. To categorize each sample, we perform naive zero-shot inference using the target MLLM and evaluate whether the model's response semantically matches the ground-truth label. If a match is observed, the sample is classified into the parameter knowledge set $D_{PK}$; otherwise, it is assigned to the world knowledge set $D_{WK}$. This process is formally defined as:

\begin{equation}
(v_i, x_i, y_i)\in
\begin{cases}
D_{PK}, & \text{if }\;\mathcal{M}_t(v_i,x_i)=y_i\\
D_{WK}, & \text{if }\;\mathcal{M}_t(v_i,x_i)\neq y_i
\end{cases},
\label{eq1:zero-shot}
\end{equation}

where $v_i$, $x_i$ and $y_i$ represent the image, query and ground-truth label in each sample, and $i=1,\dots,F$. $\mathcal{M}_t$ denotes the target MLLM.

\input{tables/experiment_results}

\noindent\textbf{$\bullet\quad$Stage 2: In-Context Learning Probing and Selection}
Taking a demonstration $(v_i,x_i,y_i)$ as part of the input, ICL guides the MLLM to understand the task and respond to the new query $(v_j,x_j)$ based on the provided context. This paradigm creates a fertile ground for exploring the interplay between different samples, i.e., whether, and to what extent, the parameter knowledge can guide the MLLM to correctly handle new world knowledge. For each sample $(v_i,x_i,y_i)$ in $D_{PK}$, we treat it as the demonstration in a one-shot ICL setup. Then we randomly select $R$ samples $v_j,x_j$ from $D_{WK}$ as query inputs, perform $R$ ICL inferences, and record the number of correct response $c_i$ for each demonstration. The computation is formalized as~\eqref{eq1} and~\eqref{eq2}:

\begin{equation}
(v_j,x_j,y_j)\;\overset{\mathrm{iid}}{\sim}\;D_{WK},\quad(j=1,\dots,R),
\label{eq1}
\end{equation}
\vspace{-0.6\baselineskip}
\begin{equation}
c_i= \sum_{j=1}^R\mathbf{1}\Bigl[M_t\bigl((v_i,x_i,y_i),(v_j,x_j)\bigr)=y_j\Bigr],
\label{eq2}
\end{equation}

where $\mathbf{1}[\cdot]$ is an indicator function that returns 1 if the response of $M_t$ semantically matches the ground-truth label $y_j$. We further introduce a threshold $\tau$ to construct $D_{ICL\_C}$, in order to analyze how varying levels of useful parametric knowledge affect coreset selection and model performance. The remaining samples, which offer no effective guidance, are excluded and form $D_{ICL\_IC}$. The process is defined as~\eqref{eq3} and~\eqref{eq4}:

\begin{equation}
D_{\mathrm{ICL\_C}} = \{(v_i,x_i,y_i)\in D_{\mathrm{PK}}\mid c_i\ge\tau\},
\label{eq3}
\end{equation}
\vspace{-0.6\baselineskip}
\begin{equation}
D_{\mathrm{ICL\_IC}} = D_{\mathrm{PK}}\setminus D_{\mathrm{ICL\_C}}.
\label{eq4}
\end{equation}

We consider $D_{ICL\_C} \cup D_{WK}$ as the preliminarily filtered coreset, where the latter is intended to be guided by the former and subsequently injected into the MLLM. To further reduce the number of training samples, we take a closer look at $D_{WK}$. A sample in $D_{WK}$ is assigned to $D_{W2C}$ if any demonstration from $D_{PK}$ enables the MLLM to predict it correctly; otherwise, it is placed in $D_{WW}$. Equation~\eqref{eq5} and~\eqref{eq6} defines this process:

\begin{equation}
\begin{split}
D_{\mathrm{W2C}} = \bigl\{(v_j,x_j,y_j)\mid\ \exists\,i\ \text{s.t.}\ 
\mathrm{M}_t\bigl((v_i,x_i,y_i), \\
\quad (v_j,x_j)\bigr) = y_j \bigr\}
\end{split}
\label{eq5}
\end{equation}
\vspace{-0.6\baselineskip}
\begin{equation}
D_{\mathrm{WW}} = D_{\mathrm{WK}}\setminus D_{\mathrm{W2C}}.
\label{eq6}
\end{equation}

Finally, we consider the four subsets $D_{ICL\_C}$, $D_{WK}$, $D_{W2C}$ and $D_{WW}$ as useful candidates for the final coreset.

%% file: tables/experiment_results.tex
\begin{table*}[!t]
\centering
\caption{Experimental results for Qwen2.5-VL-7B and Llama-3.2-11B-Vision-Instruct on VRSBench, ScienceQA, and A-OKVQA.}
\label{tab1:experiment_results}
\resizebox{0.95\textwidth}{!}{%
\begin{tabular}{lcccccc}
\Xhline{1.2pt}
\multicolumn{1}{c}{\multirow{2}{*}{Method}} & \multicolumn{2}{c}{VRSBench} & \multicolumn{2}{c}{ScienceQA} & \multicolumn{2}{c}{A-OKVQA} \\
\multicolumn{1}{c}{}             & Accuracy        & \# Samples    & Accuracy        & \# Samples    & Accuracy        & \# Samples    \\ \Xhline{1.2pt}
Qwen2.5-VL-7B                    & 57.87\%         & 0             & 80.95\%         & 0             & 77.29\%         & 0             \\
\textcolor{gray}{+ $D_{{Full}}$}                  & \textcolor{gray}{69.78\% $ \uparrow $ 11.91\%} & \textcolor{gray}{10000}         & \textcolor{gray}{96.30\% $ \uparrow $ 15.35\%} & \textcolor{gray}{5000}          & \textcolor{gray}{85.76\% $ \uparrow $ 8.47\%}  & \textcolor{gray}{5000}          \\
+ $D_{Random} $ & 67.28\% $ \downarrow $ 2.50\%  & 5000 $ \downarrow $ 50.00\% & 94.95\% $ \downarrow $ 1.35\% & 3500 $ \downarrow $ 30.00\%  & 85.52\% $ \downarrow $ 0.24\%  & 3500 $ \downarrow $ 30.00\%  \\
+ $D_{Perplexity}$~\cite{perplexity} & 67.85\% $ \downarrow $ 1.93\%  & 5000 $ \downarrow $ 50.00\% & 95.10\% $ \downarrow $ 1.20\% & 3500 $ \downarrow $ 30.00\%  & 85.50\% $ \downarrow $ 0.26\%  & 3500 $ \downarrow $ 30.00\%  \\
+ $D_{ICL\_C} $ $\cup $ $D_{WK} $ ($\tau$=1) & \textbf{70.16\%} $ \uparrow $ 0.38\%  & 7590 $ \downarrow $ 24.10\% & \underline{96.05\%} $ \downarrow $ 0.25\% & 4559 $ \downarrow $ 8.82\%  & \underline{86.81\%} $ \uparrow $ 1.05\%  & 4710 $ \downarrow $ 5.80\%  \\
+ $D_{ICL\_C} $ $\cup $ $D_{W2C} $ ($\tau$=1)   & 68.75\% $ \downarrow $ 1.03\% & 4646 $ \downarrow $ 53.54\% & 94.05\% $ \downarrow $ 2.25\% & 4205 $ \downarrow $ 15.90\% & 86.46\% $ \uparrow $ 0.70\%  & 4323 $ \downarrow $ 13.54\% \\
+ $D_{ICL\_C} $ $\cup $ $D_{{WW}} $ ($\tau$=1)      & \underline{69.55\%} $ \downarrow $ 0.23\% & 6104 $ \downarrow $ 38.96\% & 95.60\% $ \downarrow $ 0.70\% & 3908 $ \downarrow $ 21.84\% & 84.54\% $ \downarrow $ 1.22\% & 3982 $ \downarrow $ 20.36\% \\
+ $D_{ICL\_C} $ $\cup $ $D_{WK} $ ($\tau$=2) & 67.67\% $ \downarrow $ 2.11\% & 5409 $ \downarrow $ 45.91\% & \textbf{96.10\%} $ \downarrow $ 0.20\% & 3347 $ \downarrow $ 33.06\% & 85.85\% $ \uparrow $ 0.09\%  & 3814 $ \downarrow $ 23.72\% \\
+ $D_{ICL\_C} $ $\cup $ $D_{W2C} $ ($\tau$=2)   & 66.99\% $ \downarrow $ 2.79\% & 2465 $ \downarrow $ 75.35\% & 94.00\% $ \downarrow $ 2.30\% & 2993 $ \downarrow $ 40.14\% & \textbf{87.77\%} $ \uparrow $ 2.01\%  & 3427 $ \downarrow $ 31.46\% \\
+ $D_{ICL\_C} $ $\cup $ $D_{{WW}} $ ($\tau$=2)      & 65.94\% $ \downarrow $ 3.84\% & 3923 $ \downarrow $ 60.77\% & 93.90\% $ \downarrow $ 2.40\% & 2696 $ \downarrow $ 46.08\% & \underline{86.81\%} $ \uparrow $ 1.05\%  & 3086 $ \downarrow $ 38.28\% \\ \hline
Llama-3.2-11B-Vision-Instruct    & 56.52\%         & 0             & 85.95\%         & 0             & 78.52\%         & 0             \\
\textcolor{gray}{+ $D_{{Full}}$}                  & \textcolor{gray}{69.87\% $ \uparrow $ 13.35\%} & \textcolor{gray}{10000}         & \textcolor{gray}{93.80\% $ \uparrow $ 7.85\%}  & \textcolor{gray}{5000}          & \textcolor{gray}{84.19\% $ \uparrow $ 5.67\%}  & \textcolor{gray}{5000}          \\
+ $D_{Random} $ & 67.13\% $ \downarrow $ 2.74\%  & 5000 $ \downarrow $ 50.00\% & 91.18\% $ \downarrow $ 2.62\% & 3500 $ \downarrow $ 30.00\%  & 83.02\% $ \downarrow $ 1.17\%  & 3500 $ \downarrow $ 30.00\%  \\
+ $D_{Perplexity}$~\cite{perplexity} & 67.16\% $ \downarrow $ 2.71\%  & 5000 $ \downarrow $ 50.00\% & 91.25\% $ \downarrow $ 2.55\% & 3500 $ \downarrow $ 30.00\%  & 82.97\% $ \downarrow $ 1.22\%  & 3500 $ \downarrow $ 30.00\%  \\
+ $D_{ICL\_C} $ $\cup $ $D_{WK} $ ($\tau$=1) & \textbf{69.60\%} $ \downarrow $ 0.27\% & 7740 $ \downarrow $ 22.60\% & \textbf{94.35\%} $ \uparrow $ 0.55\%  & 4737 $ \downarrow $ 5.26\%  & 83.93\% $ \downarrow $ 0.26\% & 4772 $ \downarrow $ 4.56\%  \\
+ $D_{ICL\_C} $ $\cup $ $D_{W2C} $ ($\tau$=1)   & 67.20\% $ \downarrow $ 2.67\% & 4859 $ \downarrow $ 51.41\% & 92.85\% $ \downarrow $ 0.95\% & 4633 $ \downarrow $ 7.34\%  & 83.23\% $ \downarrow $ 0.96\% & 4352 $ \downarrow $ 12.96\% \\
+ $D_{ICL\_C} $ $\cup $ $D_{{WW}} $ ($\tau$=1)      & \underline{68.54\%} $ \downarrow $ 1.33\% & 6084 $ \downarrow $ 39.16\% & 92.10\% $ \downarrow $ 1.70\% & 4136 $ \downarrow $ 17.28\% & \textbf{84.63\%} $ \uparrow $ 0.44\%  & 4128 $ \downarrow $ 17.44\% \\
+ $D_{ICL\_C} $ $\cup $ $D_{WK} $ ($\tau$=2) & 68.03\% $ \downarrow $ 1.84\% & 5552 $ \downarrow $ 44.48\% & \underline{92.90\%} $ \downarrow $ 0.90\% & 3717 $ \downarrow $ 25.66\% & \underline{84.45\%} $ \uparrow $ 0.26\%  & 3851 $ \downarrow $ 22.98\% \\
+ $D_{ICL\_C} $ $\cup $ $D_{W2C} $ ($\tau$=2)   & 66.65\% $ \downarrow $ 3.22\% & 2671 $ \downarrow $ 73.29\% & 92.60\% $ \downarrow $ 1.20\% & 3613 $ \downarrow $ 27.74\% & 83.76\% $ \downarrow $ 0.43\% & 3431 $ \downarrow $ 31.38\% \\
+ $D_{ICL\_C} $ $\cup $ $D_{{WW}} $ ($\tau$=2)      & 66.60\% $ \downarrow $ 3.27\% & 3896 $ \downarrow $ 61.04\% & 91.20\% $ \downarrow $ 2.60\% & 3116 $ \downarrow $ 37.68\% & 82.53\% $ \downarrow $ 1.66\% & 3207 $ \downarrow $ 35.86\% \\ \Xhline{1.2pt}
\end{tabular}%
}
\vspace{-1.6\baselineskip}
\end{table*}

%% file: sec/4_experiments.tex
\section{Experiments}

\subsection{Experimental Setup}
\textbf{Datasets.} We evaluate CLIPPER on three datasets, each representing a distinct downstream task: (1) VRSBench~\cite{vrsbench} for remote sensing image interpretation; (2) ScienceQA~\cite{scienceqa} for science question answering; (3) A-OKVQA~\cite{a-okvqa} for outside-knowledge visual question answering. Each dataset includes training and validation (or testing) sets, from which we randomly select 10k, 5k, and 5k samples for fine-tuning, and 10k, 2k, and 1k samples for evaluation, respectively. Additionally, CHOICE~\cite{choice} benchmark is used to evaluate out-of-domain performance under different coreset settings in the remote sensing experiments.

\begin{figure*}[!t]
    \centering
    \includegraphics[width=0.85\textwidth]{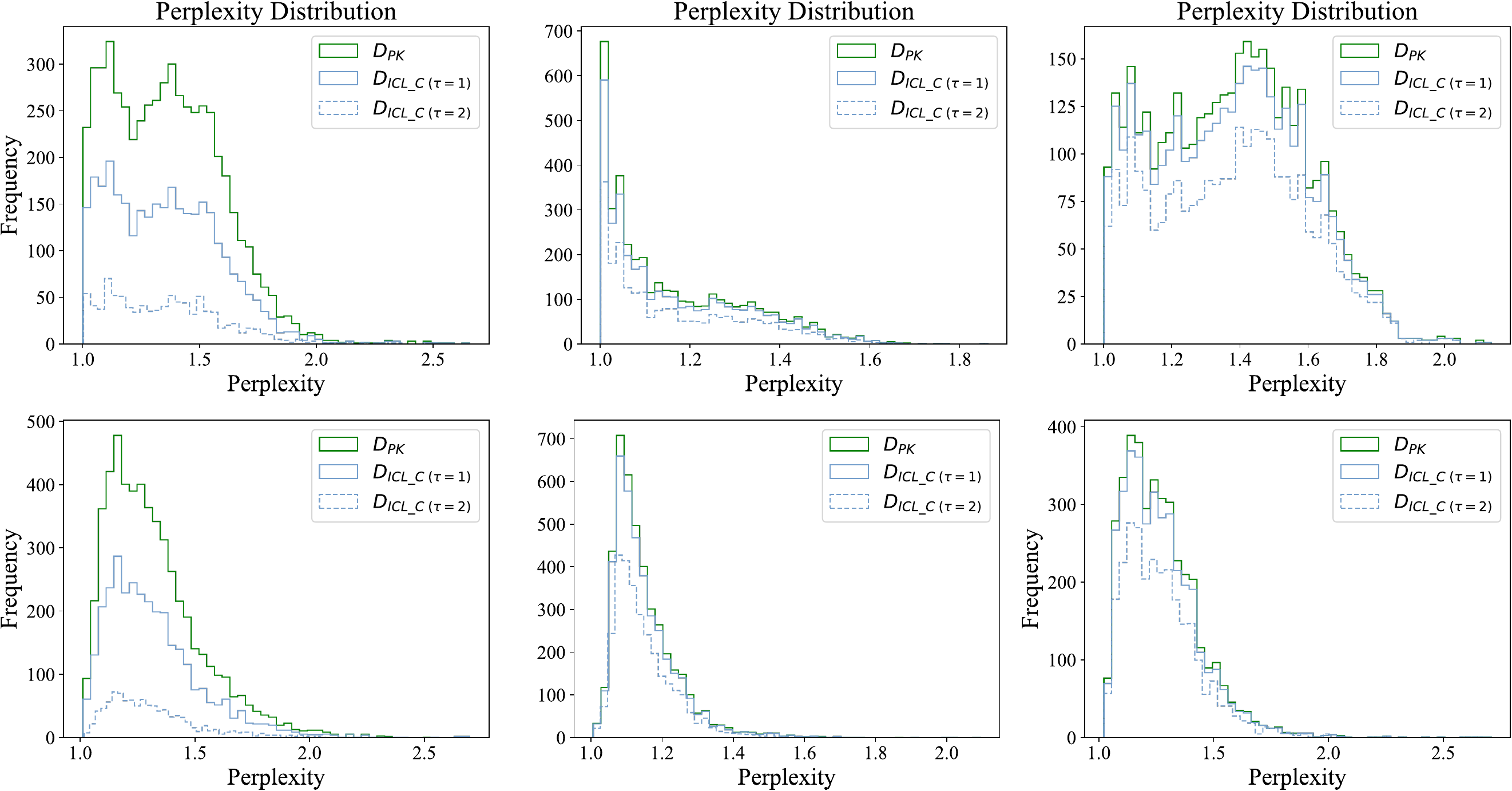}
    \caption{Perplexity distribution for Qwen2.5-VL-7B (top row) and Llama-3.2-11B-Vision-Instruct (bottom row) on VRSBench (left column), ScienceQA (middle column) and A-OKVQA (right column)} 
    \label{fig4:perplexity}
    \vspace{-1.6\baselineskip}
\end{figure*}

\textbf{Models.} The integration of ICL necessitates a large maximum context length of MLLM. For each downstream task, we evaluate the performance of two mainstream MLLMs: Qwen2.5-VL-7B~\cite{qwen2.5-vl} and Llama-3.2-11B-Vision-Instruct~\cite{llama-3.2}, both supporting a maximum context length of 128k.

\textbf{Evaluation.} All experiments are conducted on four NVIDIA 3090 GPUs using the ms-swift framework~\cite{swift}. For TSF, we perform one fully fine-tuning and six coreset-based fine-tunings: $D_{ICL\_C}\cup D_{WK}$, $D_{ICL\_C}\cup D_{W2C}$ and $D_{ICL\_C}\cup D_{WW}$, each with $\tau=1$ and $\tau=2$. We apply LoRA~\cite{lora} and train for 5 epochs on Qwen2.5-VL-7B and 3 epochs on Llama-3.2-11B-Vision-Instruct. For zero-shot and ICL inference, we use the high-throughput vLLM~\cite{vllm} framework with greedy decoding for both models, and set $R=5$. Semantic matching is conducted using the LLM-as-judge approach with GPT-4o~\cite{gpt4}, comparing the model-generated answers against the ground truth. Model performance is primarily evaluated using accuracy, with the zero-shot performance serving as the baseline. We also compare against models fine-tuned on randomly selected datasets and those selected based on middle-range perplexity distributions~\cite{perplexity}. Additionally, total elapsed time is recorded for all experiments.

\subsection{Main Results and Key Findings}
The experimental results are presented in Table~\ref{tab1:experiment_results}, comparing the performance of the baseline, full fine-tuning and different combinations of the four subsets selected by CLIPPER, along with the corresponding number of training samples used in each setting (Complete experimental results are presented in Tables~\ref{tab:time_spent} and \ref{tab:choice_results} in the Appendix.).

\noindent\textbf{$\bullet\quad \mathbf{D_{ICL\_C}\cup D_{WK}}$ Consistently Matches Full Fine-tuning with Fewer Samples.} Quantitative results in Table~\ref{tab1:experiment_results} show that finetuning on the combination of $D_{ICL\_C}$ and $D_{WK}$ yields performance comparable to full fine-tuning across both MLLMs. This observation highlights the significant redundancy in training samples that overlap with parameter knowledge of target MLLM, and demonstrates that selectively identifying samples capable of guiding world knowledge through ICL is an effective strategy for TSF data selection. For instance, Qwen2.5-VL-7B even outperforms the fully fine-tuned counterpart by 0.38\% on VRSBench while using 24.1\% fewer samples. A similar trend holds for LLaMA-3.2-11B-Vision-Instruct, where removing 22.6\% of redundant samples incurs only a 0.27\% performance drop. When a stricter threshold $\tau = 2$ is applied, further filtering $D_{ICL_C}$ for stronger guidance capability, the training data shrinks more aggressively. For LLaMA-3.2-11B-Vision-Instruct, data is reduced by 44.48\% and 25.66\% on VRSBench and ScienceQA, with only 1.84\% and 0.9\% performance decreases, respectively..

\noindent\textbf{$\bullet\quad$Aggressive Pruning of $\mathbf{D_{WK}}$ Achieves Major Sample Reduction with Minor Performance Drop.} Building on the effectiveness of $D_{ICL\_C}$, we further investigate the individual influence of $D_{W2C}$ and $D_{WW}$, which correspond to world knowledge samples that the MLLM can correctly respond to via ICL and those it consistently fails to handle, respectively. Results show that removing either subset causes only a slight performance decrease relative to using the full $D_{WK}$, while yielding substantial reductions in training data, particularly under larger $\tau$ values. For example, on VRSBench, Qwen2.5-VL-7B achieves a 75.35\% reduction in training data by excluding $D_{WW}$, with only a further 0.68\% performance drop. Interestingly, on A-OKVQA, only including $D_{W2C}$ contributes to a 31.46\% reduction in training data while yielding a 1.92\% increase in accuracy.

\noindent\textbf{$\bullet\quad$CLIPPER Maintains Out-of-domain Generalization.} A key question in data selection is whether the selected training data benefits only the source dataset or also transfers to unseen domains. To evaluate this, we assess the generalization capability of the fine-tuned models on VRSBench by testing them on the out-of-domain remote sensing benchmark CHOICE. As shown in Table~\ref{tab:choice_results}, LLaMA-3.2-11B-Vision-Instruct experiences at most a 2\% performance drop, while all versions of Qwen2.5-VL-7B show consistent performance improvements. These results demonstrate that the coresets selected by CLIPPER are not only representative for in-domain learning but also robust in out-of-domain generalization.

\noindent\textbf{$\bullet\quad$CLIPPER Takes Effect Through Perplexity-based Selection.} To understand the type of data favored by CLIPPER for each downstream task, we analyze the perplexity distribution of all subsets selected by CLIPPER in Fig.~\ref{fig4:perplexity}. $D_{ICL\_C}$ (for both $\tau=1$ and $\tau=2$) exhibits a perplexity distribution closely aligned with that of $D_{PK}$. This suggests the interplay between parameter knowledge and world knowledge in ICL tends to select samples with strong guidance capability, and that these samples effectively balance perplexity-based importance and diversity. These characteristics contribute to CLIPPER's stable and competitive performance while significantly reducing computational costs. Moreover, the inferior performance of the middle-range perplexity selection highlights that a single heuristic metric tends to ignore representative samples assigned within low-score regions. In contrast, the similarity between $D_{W2C}$ or $D_{WW}$ and $D_{WK}$ 
do not consistently correlate with improved performance (See detailed analysis in Appendix). We attribute this to differences in architecture design, inherent parameter knowledge, and multi-modal alignment strategies, which can lead to distinct internal representations of world knowledge.

%% file: sec/5_conclusion.tex
\section{Conclusion}
In this paper, we investigate the underexplored interplay between parameter knowledge and world knowledge embedded in training samples during TSF. Given the additional fine-tuning costs and the limited focus on either data importance or data diversity in existing data selection methods, we propose CLIPPER, a training-free pipeline that leverages zero-shot inference and ICL probing for efficient data selection.
Experiments on two MLLMs and three downstream tasks demonstrate that CLIPPER significantly reduces training samples while maintaining performance on par with full fine-tuning. Moreover, the ICL-based probing enables perplexity-aware data selection, as evidenced by the highly similar perplexity distributions between selected subsets and original datasets. CLIPPER inspires broader use of ICL in data selection and facilitates efficient MLLM deployment in domain-specific applications.

%% file: sec/X_appendix.tex
\clearpage
\section*{Appendix}
\label{appendix}

\subsection{Algorithm}
The algorithm of CLIPPER is presented in Algorithm~\ref{alg:CLIPPER}.
\input{tables/algorithm}

\subsection{Datasets Details}
We evaluate CLIPPER on three datasets, each representing a distinct downstream task: (1) VRSBench~\cite{vrsbench} for remote sensing image interpretation; (2) ScienceQA~\cite{scienceqa} for science question answering; (3) A-OKVQA~\cite{a-okvqa} for outside-knowledge visual question answering. Additionally, we use CHOICE~\cite{choice} benchmark to evaluate out-of-domain performance under different coreset settings in the remote sensing experiments. For our experiments, we focus on the VQA task, which consists of 85,813 training samples and 37,409 testing samples. We randomly select 10,000 samples for both training and testing for TSF. In ScienceQA, we exclude text-only and image-only QA pairs, randomly selecting 5,000 samples from the 6,218 training samples for TSF, and 2,000 samples from the 2,017 testing samples. For the A-OKVQA dataset, we randomly select 5,000 samples from the 17,056 training samples and use all 1,145 samples from the validation set for testing. For the CHOICE benchmark, we specifically focus on the tasks within the Perception dimension for remote sensing image interpretation.

\subsection{Interplay among Training Samples}
We conduct the first experiment exploring the interplay among training samples in the context of multi-modality, as illustrated in Fig.~\ref{fig2:fraction}. For this experiment, we use 10,000 training samples from VRSBench and the Qwen2.5-VL-7B model. Initially, we perform zero-shot inference to categorize the samples into parameter knowledge and world knowledge, resulting in 5,570 and 4,430 samples, respectively. We then fix the number of fine-tuning samples to 4,430 and vary the proportions of these two types of samples to perform TSF on the Qwen2.5-VL-7B model. The proportions of parameter knowledge are set at 0\%, 25\%, 50\%, 75\%, and 100\%. Finally, we evaluate the fine-tuned models on a test set of 10,000 samples, and present the quantitative results in Table.

\input{tables/fraction}

\subsection{More Experimental Results}
Total elapsed time is recorded for all experiments in Table~\ref{tab:time_spent} and the out-of-domain evaluation on the CHOICE benchmark is presented in Table~\ref{tab:choice_results}.

\input{tables/time_spent}

\subsection{Detailed Analysis}
\begin{figure*}[!htb]
    \centering
    \includegraphics[width=1.0\textwidth]{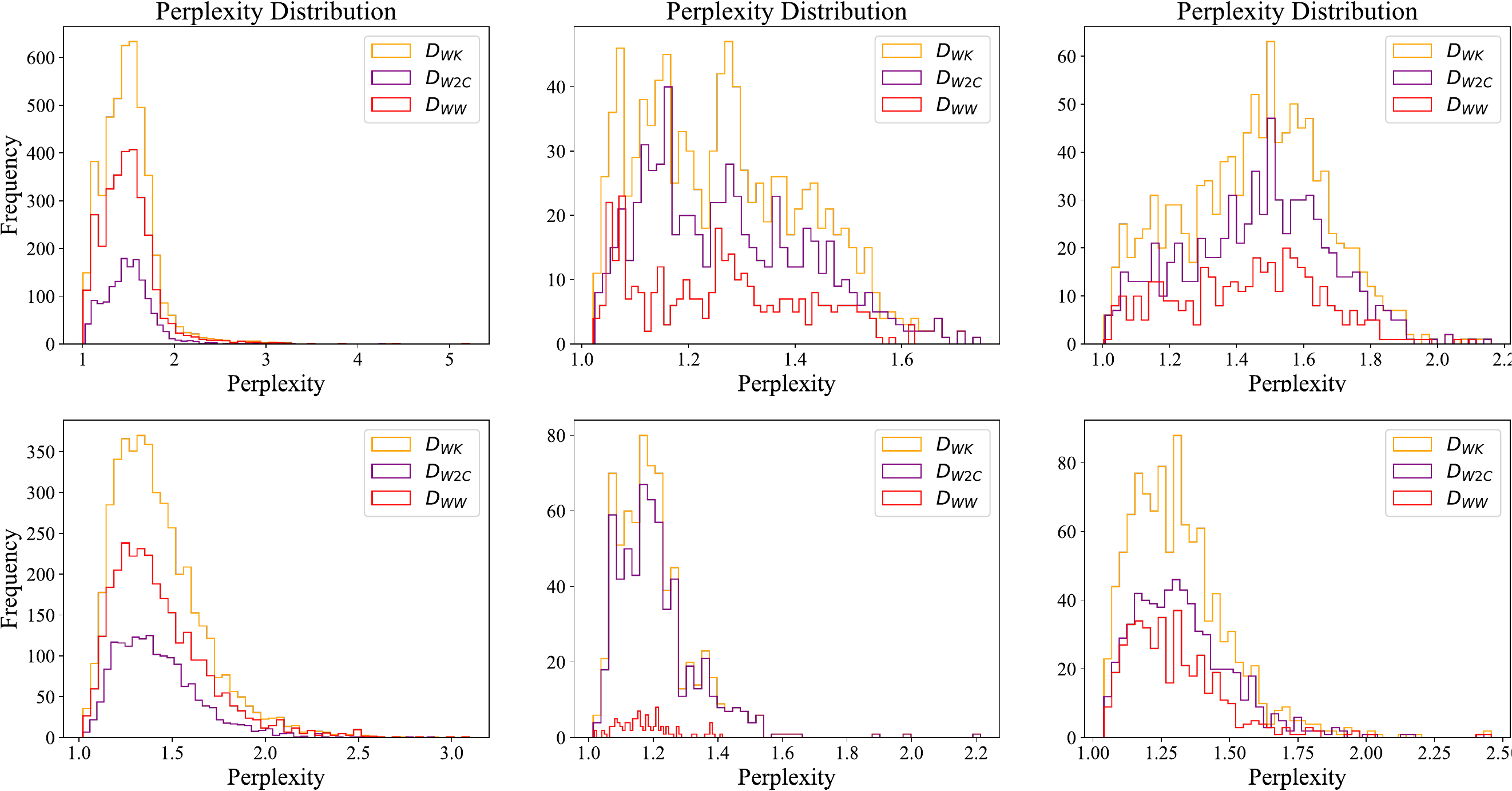}
    \caption{Perplexity distribution for Qwen2.5-VL-7B (top row) and Llama-3.2-11B-Vision-Instruct (bottom row) on VRSBench (left column), ScienceQA (middle column) and A-OKVQA (right column)} 
    \label{fig5:perplexity}
\end{figure*}

We analyze the perplexity distribution of $D_{W2C}$ or $D_{WW}$ selected by CLIPPER in Fig.~\ref{fig5:perplexity}. The similarity between $D_{W2C}$ or $D_{WW}$ and $D_{WK}$ do not consistently correlate with improved performance. Therefore, the aggressive pruning of $D_{WK}$ needs further consideration in practice. For VRSBench, the perplexity distribution of $D_{WW}$ more closely resembles that of the original $D_{WK}$ across both MLLMs. Accordingly, a similar trend is observed: when $\tau=1$, the performance of $D_{ICL_C} \cup D_{WW}$ surpasses that of $D_{ICL_C} \cup D_{WW}$, suggesting that world knowledge samples that can be corrected are less beneficial than those that cannot. However, this trend reverses when $\tau = 2$. In contrast, on ScienceQA, it is $D_{W2C}$ that exhibits a perplexity distribution more similar to $D_{WK}$ across both MLLMs. Despite this, Qwen2.5-VL-7B shows a trend similar to that observed on VRSBench: $D_{ICL_C} \cup D_{WW}$ outperforms $D_{ICL_C} \cup D_{W2C}$ by 1.55\% when $\tau=1$, but slightly underperforms by 0.1\% when $\tau=2$. Interestingly, LLaMA-3.2-11B-Vision-Instruct shows a contrasting performance trend compared to Qwen2.5-VL-7B. For A-OKVQA, the perplexity distribution of $D_{W2C}$ in Qwen2.5-VL-7B closely mirrors that of $D_{WK}$, while for LLaMA-3.2-11B-Vision-Instruct, it is $D_{WW}$ that aligns more closely with $D_{WK}$. As a result, Qwen2.5-VL-7B consistently shows that $D_{ICL_C} \cup D_{W2C}$ outperforms $D_{ICL_C} \cup D_{WW}$ for both $\tau=1$ and $\tau=2$. However, LLaMA-3.2-11B-Vision-Instruct continues to behave differently. We attribute these discrepancies to differences in architectural design, the amount and nature of encoded world knowledge, and the models’ multi-modal alignment strategies, which may result in distinct internal representations of world knowledge. Nonetheless, these observations warrant further investigation.

\input{tables/CHOICE_results}

%% file: tables/algorithm.tex
\begin{algorithm}[!b]
\caption{CLIPPER for Data Selection}
\label{alg:CLIPPER}

\textbf{Input}:
\begin{tabular}[t]{@{}l @{\,--- } l@{}}
  $D_{\mathrm{Full}}$ & the whole TSF dataset $\{(v_f,x_f,y_f)\}_{f=1}^{F}$; \\
  $M_t$ & Target MLLM to be fine-tuned; \\
  $\tau$ & Threshold for $D_{\mathrm{ICL\_C}}$; \\
  $R$ & Number of ICL for each sample in $D_{\mathrm{PK}}$; \\
\end{tabular}


\begin{algorithmic}[1] 
\STATE \textbf{procedure} CLIPPERSelection ($D_{\mathrm{Full}}$, $M_t$, $\tau$, $R$)
\FOR{each $\{(v_f,x_f,y_f)\}$ in $D_{\mathrm{Full}}$}
\IF {$y_f \neq \mathrm{M}_t(v_f,x_f)$}
\STATE $D_{\mathrm{WK}} \leftarrow D_{\mathrm{WK}} \cup \{(v_f,x_f,y_f)\}$
\ELSE
\STATE $D_{\mathrm{PK}} \leftarrow D_{\mathrm{PK}} \cup \{(v_f,x_f,y_f)\}$
\ENDIF
\ENDFOR


\FOR{each $\{(v_i,x_i,y_i)\}$ in $D_{\mathrm{PK}}$}
\STATE $c \leftarrow 0$
\FOR {$j$ from $1$ to $R$}
\STATE $(v_j,x_j,y_j)\leftarrow \mathrm{RandomSelect}(D_{PK})$
\IF{$\mathrm{M}_t((v_i,x_i,y_i),(v_j,x_j)) == y_j$}
\STATE $c \leftarrow c + 1$
\STATE $D_{\mathrm{W2C}} \leftarrow D_{\mathrm{W2C}} \cup \{(v_j,x_j,y_j)\}$
\ENDIF
\ENDFOR
\IF{$c \geq \tau$}
\STATE $D_{\mathrm{ICL\_C}} \leftarrow D_{\mathrm{ICL\_C}} \cup \{(v_i,x_i,y_i)\}$
\ENDIF
\ENDFOR
\STATE $D_{\mathrm{ICL\_IC}} \leftarrow D_{\mathrm{PK}} - D_{\mathrm{ICL\_C}}$
\STATE $D_{\mathrm{WW}} \leftarrow D_{\mathrm{WK}} - D_{\mathrm{W2C}}$
\STATE \textbf{return} $D_{\mathrm{ICL\_C}}$, $D_{\mathrm{ICL\_IC}}$, $D_{\mathrm{W2C}}$, $D_{\mathrm{WW}}$
\end{algorithmic}
\end{algorithm}

%% file: tables/fraction.tex
\begin{table}[!t]
\centering
\caption{Results of Qwen2.5-VL-7B fine-tuned on the VRSBench training set, using varying proportions of parameter knowledge and world knowledge, while keeping the total number of training samples fixed. }
\label{tab3:fraction}
\resizebox{\linewidth}{!}{
\begin{tabular}{ccccccc}
\Xhline{1.2pt}
           & \multicolumn{6}{c}{Accuracy}                                                     \\
\multirow{-2}{*}{Model} & {\color[HTML]{808080} Baseline} & Epoch-1 & Epoch-2 & Epoch-3 & Epoch-4 & Epoch-5 \\ \Xhline{1.2pt}
P-0\_W-100 & {\color[HTML]{808080} 57.87\%} & 64.72\% & 68.10\% & 66.43\% & 66.83\% & 67.01\% \\
P-25\_W-75 & {\color[HTML]{808080} 57.87\%} & 68.78\% & 69.40\% & 69.44\% & 69.05\% & 67.63\% \\
P-50\_W-50 & {\color[HTML]{808080} 57.87\%} & 69.44\% & 69.82\% & 69.72\% & 69.17\% & 68.25\% \\
P-75\_W-25 & {\color[HTML]{808080} 57.87\%} & 68.27\% & 70.13\% & 69.25\% & 68.57\% & 68.29\% \\
P-100\_W-0 & {\color[HTML]{808080} 57.87\%} & 66.40\% & 66.89\% & 66.89\% & 66.53\% & 66.42\% \\ \Xhline{1.2pt}
\end{tabular}
}
\end{table}

%% file: tables/time_spent.tex
\begin{table}[!b]
\centering
\caption{Comparison of fine-tuning time (in seconds, including two-stage inference when CLIPPER is applied) for Qwen2.5-VL-7B and Llama-3.2-11B-Vision-Instruct on VRSBench, ScienceQA, and A-OKVQA.}
\label{tab:time_spent}
\resizebox{\columnwidth}{!}{%
\begin{tabular}{lccc}
\Xhline{1.2pt}
\multicolumn{1}{c}{Method}         & VRSBench   & ScienceQA   & A-OKVQA   \\ \Xhline{1.2pt}
\multicolumn{4}{c}{\cellcolor[HTML]{E8E8E8}Qwen2.5-VL-7B}                 \\
\textcolor{gray}{+ $D_{Full}$}                    & \textcolor{gray}{52841}      & \textcolor{gray}{23105}       & \textcolor{gray}{20184}     \\
+ $D_{ICL\_C}$ $\cup$ $D_{WK}$ ($\tau$=1)   & 44928      & 22296       & 20140     \\
+ $D_{ICL\_C}$ $\cup$ $D_{W2C}$ ($\tau$=1)     & 28183      & 20339       & 18576     \\
+ $D_{ICL\_C}$ $\cup$ $D_{WW}$ ($\tau$=1)        & 36144      & 19587       & 17090     \\
+ $D_{ICL\_C}$ $\cup$ $D_{WK}$ ($\tau$=2)   & 30012      & 16125       & 16265     \\
+ $D_{ICL\_C}$ $\cup$ $D_{W2C}$ ($\tau$=2)     & 14499      & 14522       & 14737     \\
+ $D_{ICL\_C}$ $\cup$ $D_{WW}$ ($\tau$=2)        & 21977      & 13726       & 13299     \\ \hline
\multicolumn{4}{c}{\cellcolor[HTML]{E8E8E8}Llama-3.2-11B-Vision-Instruct} \\
\textcolor{gray}{+ $D_{Full}$}                    & \textcolor{gray}{92642}      & \textcolor{gray}{48747}       & \textcolor{gray}{47477}     \\
+ $D_{ICL\_C}$ $\cup$ $D_{WK}$ ($\tau$=1)   & 74999      & 45978       & 46311     \\
+ $D_{ICL\_C}$ $\cup$ $D_{W2C}$ ($\tau$=1)     & 47037      & 45716       & 41640     \\
+ $D_{ICL\_C}$ $\cup$ $D_{WW}$ ($\tau$=1)        & 59409      & 40973       & 39568     \\
+ $D_{ICL\_C}$ $\cup$ $D_{WK}$ ($\tau$=2)   & 54475      & 37286       & 36846     \\
+ $D_{ICL\_C}$ $\cup$ $D_{W2C}$ ($\tau$=2)     & 26767      & 35831       & 33865     \\
+ $D_{ICL\_C}$ $\cup$ $D_{WW}$ ($\tau$=2)        & 38213      & 30803       & 31208     \\ \Xhline{1.2pt}
\end{tabular}%
}
\end{table}

%% file: tables/CHOICE_results.tex
\begin{table}[!t]
\centering
\caption{Experimental results of Qwen2.5-VL-7B and Llama-3.2-11B-Vision-Instruct evaluated on the CHOICE subset after fine-tuning on VRSBench.}
\label{tab:choice_results}
\resizebox{0.9\columnwidth}{!}{%
\begin{tabular}{lcc}
\Xhline{1.2pt}
\multicolumn{1}{c}{Method}       & \multicolumn{2}{c}{CHOICE (Subset)} \\ \Xhline{1.2pt}
Qwen2.5-VL-7B                    & \multicolumn{2}{c}{77\%}            \\
\textcolor{gray}{+ $D_{Full}$}                  & \textcolor{gray}{77\%}             & \textcolor{gray}{-}                \\
+ $D_{ICL\_C}$ $\cup$ $D_{WK}$   ($\tau$=1) & 78\%             & $ \uparrow $1\%             \\
+ $D_{ICL\_C}$ $\cup$ $D_{W2C}$   ($\tau$=1)   & 78\%             & $ \uparrow $1\%             \\
+ $D_{ICL\_C}$ $\cup$ $D_{WW}$ ($\tau$=1)      & 79\%             & $ \uparrow $2\%             \\
+ $D_{ICL\_C}$ $\cup$ $D_{WK}$   ($\tau$=2) & 78\%             & $ \uparrow $1\%             \\
+ $D_{ICL\_C}$ $\cup$ $D_{W2C}$   ($\tau$=2)   & 79\%             & $ \uparrow $2\%             \\
+ $D_{ICL\_C}$ $\cup$ $D_{WW}$ ($\tau$=2)      & 78\%             & $ \uparrow $1\%             \\ \hline
Llama-3.2-11B-Vision-Instruct    & \multicolumn{2}{c}{70\%}            \\
\textcolor{gray}{+ $D_{Full}$}                  & \textcolor{gray}{72\%}             & \textcolor{gray}{$ \uparrow $2\%}             \\
+ $D_{ICL\_C}$ $\cup$ $D_{WK}$   ($\tau$=1) & 70\%             & $ \uparrow $2\%             \\
+ $D_{ICL\_C}$ $\cup$ $D_{W2C}$   ($\tau$=1)   & 74\%             & $ \uparrow $2\%             \\
+ $D_{ICL\_C}$ $\cup$ $D_{WW}$ ($\tau$=1)      & 72\%             & -                \\
+ $D_{ICL\_C}$ $\cup$ $D_{WK}$   ($\tau$=2) & 71\%             & $ \uparrow $1\%             \\
+ $D_{ICL\_C}$ $\cup$ $D_{W2C}$   ($\tau$=2)   & 72\%             & -                \\
+ $D_{ICL\_C}$ $\cup$ $D_{WW}$ ($\tau$=2)      & 71\%             & $ \uparrow $1\%             \\ \Xhline{1.2pt}
\end{tabular}%
}
\end{table}